\documentclass[aps,prl,twocolumn,groupedaddress,floatfix]{revtex4-1}
\usepackage{tabulary,graphicx,amsmath,amsfonts,amssymb}
\usepackage[utf8x]{inputenc}
\usepackage{lineno} 
\usepackage{mathtools}
\usepackage{xcolor}
\usepackage{subcaption}
\definecolor{blueryb}{rgb}{0.01, 0.28, 1.0}

\definecolor{greenryb}{rgb}{0.1, 1.0, 0.1}

\makeatletter
\renewenvironment{figure}
               {\@float{figure}}
               {\end@float}
\renewenvironment{figure*}
               {\@dblfloat{figure}}
               {\end@dblfloat}

\renewenvironment{table*}
               {\@dblfloat{table}}
               {\end@dblfloat}

\makeatother

\usepackage{caption}
\usepackage{url,multirow,morefloats,floatflt,cancel}
\makeatletter

\AtBeginDocument{\@ifpackageloaded{textcomp}{}{\usepackage{textcomp}}}
\makeatother
\usepackage{colortbl}
\usepackage{xcolor}
\usepackage{pifont}
\usepackage[nointegrals]{wasysym}
\makeatletter

\def\mcWidth#1{\csname TY@F#1\endcsname+\tabcolsep}

\def\cAlignHack{\rightskip\@flushglue\leftskip\@flushglue\parindent\z@\parfillskip\z@skip}
\def\rAlignHack{\rightskip\z@skip\leftskip\@flushglue \parindent\z@\parfillskip\z@skip}

\if@twocolumn\usepackage{dblfloatfix}\fi 
\AtBeginDocument{
\expandafter\ifx\csname eqalign\endcsname\relax
\def\eqalign#1{\null\vcenter{\def\\{\cr}\openup\jot\m@th
  \ialign{\strut$\displaystyle{##}$\hfil&$\displaystyle{{}##}$\hfil
      \crcr#1\crcr}}\,}
\fi
}

\AtBeginDocument{%
  \@ifpackageloaded{endfloat}%
   {\renewcommand\efloat@iwrite[1]{\immediate\expandafter\protected@write\csname efloat@post#1\endcsname{}}}{}%
}%

\let\lt=<
\let\gt=>
\def\processVert{\ifmmode|\else\textbar\fi}

\@ifundefined{subparagraph}{
\def\subparagraph{\@startsection{paragraph}{5}{2\parindent}{0ex plus 0.1ex minus 0.1ex}%
{0ex}{\normalfont\small\itshape}}%
}{}

\newcommand\role[1]{\unskip}
\newcommand\aucollab[1]{\unskip}
  
\@ifundefined{tsGraphicsScaleX}{\gdef\tsGraphicsScaleX{1}}{}
\@ifundefined{tsGraphicsScaleY}{\gdef\tsGraphicsScaleY{.9}}{}
\def\checkGraphicsWidth{\ifdim\Gin@nat@width>\linewidth
	\tsGraphicsScaleX\linewidth\else\Gin@nat@width\fi}

\def\checkGraphicsHeight{\ifdim\Gin@nat@height>.9\textheight
	\tsGraphicsScaleY\textheight\else\Gin@nat@height\fi}

\def\fixFloatSize#1{}
\let\ts@includegraphics\includegraphics

\def\inlinegraphic[#1]#2{{\edef\@tempa{#1}\edef\baseline@shift{\ifx\@tempa\@empty0\else#1\fi}\edef\tempZ{\the\numexpr(\numexpr(\baseline@shift*\f@size/100))}\protect\raisebox{\tempZ pt}{\ts@includegraphics{#2}}}}

\AtBeginDocument{\def\includegraphics{\@ifnextchar[{\ts@includegraphics}{\ts@includegraphics[width=\checkGraphicsWidth,height=\checkGraphicsHeight,keepaspectratio]}}}

\def\URL#1#2{\@ifundefined{href}{#2}{\href{#1}{#2}}}

\def\UrlOrds{\do\*\do\-\do\~\do\'\do\"\do\-}%
\g@addto@macro{\UrlBreaks}{\UrlOrds}
\makeatother

\begin{document}

\title{Isotope quantum effects in the metallization transition in liquid hydrogen}
\author{Sebastiaan van de Bund,$^1$ Heather Wiebe$^1$ \& Graeme J. Ackland$^1$}

\affiliation{
 School of Physics \& Astronomy, The University of Edinburgh, Edinburgh, EH9 3FD, United Kingdom.
}

\begin{abstract}
    Quantum effects in condensed matter normally only occur at low temperatures.
    Here we show a large quantum effect in high-pressure liquid hydrogen at thousands of Kelvins.  We show that the metallization transition in hydrogen is subject to a very large isotope effect, occurring hundreds of degrees lower than the equivalent transition in deuterium.  We examined this using path integral molecular dynamics simulations which identify a liquid-liquid transition involving atomization, metallization, and changes in viscosity,  specific heat and compressibility.  The difference between H$_2$ and D$_2$ is a quantum mechanical effect which can be associated with the larger zero-point energy in H$_2$ weakening the covalent bond.  Our results mean that experimental results on deuterium must be corrected before they are relevant to understanding hydrogen at planetary conditions.
\end{abstract}
\date{\today} 
\maketitle 



\iftrue
Hydrogen, despite being the simplest element on the periodic table, exhibits rich physics at high pressures. 
Of particular interest is the insulator-to-metal transition, wherein the system transforms from an insulating molecular phase to a conducting phase. 
Extremely high static compression is required to reach the metallic solid\cite{SilveraMetalScience,Loubeyre2020Synchotron}, but transition to a metallic liquid has been observed in both static\cite{dzyabura2013evidence,zaghoo2016evidence,Zaghoo2018Striking,Jiang2020Spectroscopic} and dynamic\cite{weir1996metallization,knudson2015direct,Celliers2018Insulator} compression experiments. 
This liquid-liquid phase transition (LLPT) is of vital importance to the modelling of the interior of Jovian-like planets\cite{becker2010modeling,swift2011mass,chabrier2019new}, as the metallization of hydrogen is thought to cause the demixing of hydrogen and helium at pressure\cite{mcmahon2012properties}.  
Despite this importance, the nature of the LLPT is not fully understood.
The prevailing hypothesis is that the LLPT is a first order transition between an insulating molecular liquid and a conducting atomic one, terminating at a critical point between 1000 and 1500 K.
This is supported by density functional theory (DFT)\cite{scandolo2003liquid,holst2008thermophysical,lorenzen2010first,morales2013nuclear,Geng2019Thermodynamic} and quantum Monte Carlo\cite{morales2010evidence,mazzola2015distinct,Pierleoni2016Liquid,Rillo2019Optical} simulations. However, a quantitative agreement between theory and experiment has not yet been achieved, and the consensus in modelling was recently challenged\cite{cheng2020evidence}.
\fi

The hydrogen phase diagram exhibits features which are impossible in classical physics.
Such features are typically only seen at low temperatures and involve nuclear quantum effects (NQE).
One example is the solid "Phase I" of hydrogen, which extends to zero temperature but involves freely-rotating molecules. 
Then there is a significant difference between H$_2$ and D$_2$ in the low temperature phase boundary between Phase I and the broken-symmetry Phase II, in which both isotope mass and quantum spin statistics play a role\cite{liu2020counterintuitive}.

However, "Low-temperature" begs the question "Low compared to what?"
No isotope effect is known for the melting line in hydrogen, and nuclear quantum effects are often ignored at the kiloKelvin temperatures of liquid hydrogen.
The melting transition involves changes in intermolecular bonding, which is a relatively weak interaction.
The LLPT, on the other hand, involves breaking the covalent bonds. 
The zero-point energy of H$_2$ is 78.46 meV ($\sim$ 910 K) higher than that of D$_2$,\cite{Irikura2007ZPE} and so the covalent bonding in H$_2$ is significantly weaker. Consequently, isotope effects at an unprecedentedly high temperature could be manifested in the experimentally observable difference in the LLPT phase boundary. 
Indeed, a 700 K difference in the LLPT phase lines of H$_2$ and D$_2$ was reported by Zaghoo \textit{et al.} by monitoring reflectivity in a laser-heated diamond anvil cell\cite{Zaghoo2018Striking}. Interestingly, a spectroscopic study conducted by Jiang \textit{et al.} one year later did not detect an appreciable isotope effect\cite{Jiang2020Spectroscopic}.

There are numerous pitfalls in calculating the hydrogen LLPT (see Supplementary Material for further discussion\cite{SMref}).
Cheng {\it et al.} recently demonstrated how many previous studies with less than 250 atoms and trajectories several ps in length were stuck in a solid rather than molecular-liquid phase.  $\Gamma$-only k-point sampling leads to unphysical chain-like "polyhydrogen" structures \cite{magduau2017theory} and inclusion of NQE\cite{morales2010equation,morales2010evidence,Pierleoni2016Liquid,Rillo2019Optical} is essential to finding any isotope effect because in classical theory, the phase boundaries of hydrogen and deuterium are identical\cite{ackland2015appraisal}.  Such a difference undermines the relevance of experiments on deuterium to determine the hydrogen equation of state - a crucial component of planetary and exoplanetary science.
 
Another challenge for DFT is that most exchange-correlation functionals cannot describe both molecular and metallic phases well, further contributing to the wide spread of results for the LLPT line from DFT  (see Table S1 in Supplementary Material\cite{SMref}).  Functionals which do not correctly describe the high density limit of the exchange energy are particularly poor for describing molecule-atom transitions\cite{azadi2017role}. Here we adopt the BLYP functional for the majority of our results\cite{blyp-B,blyp-LYP} which has been shown to be the closest to Quantum Monte Carlo for molecular systems, including molecular metals which are problematic with PBE\cite{clay2014benchmarking}.
Nevertheless, by comparing to additional simulations performed using PBE\cite{PBE} and vdw-DF\cite{dion2004van} we also show that while the location of the phase boundary strongly depends on the functional, the isotope effect is not nearly as sensitive.

 In this paper we use the ring polymer path integral molecular dynamics (PIMD) technique\cite{Marx2009AbInitio} in combination with DFT to calculate the LLPT boundary for classical H$_2$ and quantum H$_2$/D$_2$. 
 We demonstrate the large isotopic shift by monitoring various properties of the system along five isotherms ranging from 1000 to 2500 K for both H$_2$ and D$_2$.  This isotope shift is impossible in classical physics, and is conclusive evidence of important NQEs at kiloKelvin temperatures.
 


\begin{figure}[!htbp]
    \centering
    \includegraphics[width=\columnwidth]{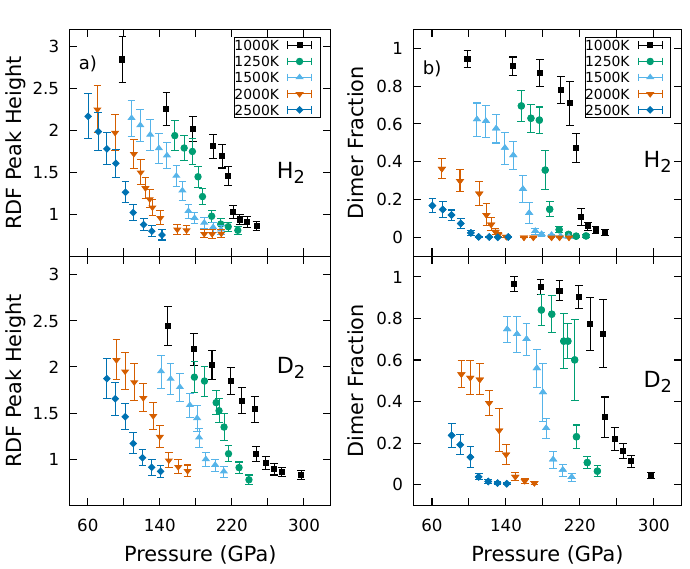}
    \caption{Evidence of molecular dissociation. (a) Height of first (normalized) RDF peak, which corresponds to the molecular bond length. While the peak becomes relatively small, it does not disappear even in the atomic phase. (b) Fractions of H$_2$ and D$_2$ dimers present, where a stable dimer is defined as two H or D atoms that are less than 0.9\AA\, apart for at least 85 fs.  Example RDFs are shown in the supplementary materials.}
    \label{fig:dissociation}
\end{figure}

One measure of the LLPT is the fraction of H$_2$ molecules present in the system. This can be obtained by tracking the height of the first peak in the radial distribution function (RDF), which corresponds to the molecular bond length. 
The results in Fig. \ref{fig:dissociation}a) clearly show dissociation occurring for both H$_2$ and D$_2$, as evidenced by a drop in peak height with increasing pressure. 
The relative sharpness and magnitude of this drop are temperature-dependent, with the low temperature isotherms exhibiting the sharpest, largest drop and high temperature isotherms exhibiting smooth, small changes in peak height.
The isotope effect is observed here as a shift in the phase boundary between H$_2$ and D$_2$.
This shift is on the order of 250 K. 



The change in RDF provides clear evidence of a transition, but does not explicitly consider the existence of molecules and does not distinguish between possible H$_n$ clusters for $n$ higher than 2. 
A more intuitive description of the dissociation can be found by considering the fraction of atoms that form a molecule. 
This is extracted from the interatomic distances, noting that the molecules continually break apart and reform, akin to a chemical reaction $\text{H}_2 \rightleftharpoons 2\text{H}$, which naturally introduces the idea of a molecular lifetime. 
We define a molecule to be two hydrogen or deuterium atoms less than 0.9 {\AA} apart for at least 85 fs. This allows for at least 10 vibrations in the case of hydrogen. 
This choice is motivated by the limits of experimental detectability: it corresponds to a spectroscopic natural linewidth of 400 cm$^{-1}$, 


The resulting dimer fraction across the PIMD runs is shown in Fig. \ref{fig:dissociation}b). 
The limiting cases of high and low temperatures show that the dimer fraction tends to the expected values of 0 and 1 in the atomic and molecular limits, although pressures near the transition here will have a sizeable fraction of the liquid already dissociated.
Like the RDF peak height, the dimer fraction drops steeply across the transition pressure for all isotherms. 
However, this drop can only be considered to be discontinuous around 1000 K for hydrogen and up to around 1250 K for deuterium; at higher temperatures the transitions become a crossover.
This suggests that the critical point of the LLPT likely lies between 1000 and 1500 K for both isotopes, and is higher in deuterium.

\begin{figure}[!htbp]
    \centering
    \includegraphics[width=0.7\columnwidth]{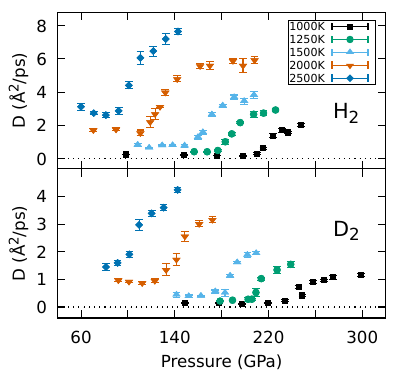}
    \caption{Diffusion constant $D$ as calculated from mean square displacements for H$_2$ and D$_2$, showing evidence of increasing diffusivity across the dissociation with a clear isotope effect.}
    \label{fig:diffusion}
\end{figure}{}

The onset of dissociation is also marked by an increase in the diffusivity. 
Fig. \ref{fig:diffusion} shows calculated diffusion constants $D$ for all isotherms, where it can be seen that $D$ markedly increases in the high pressure atomic phase. 
The curves are qualitatively similar, allowing for the two-times larger deuterium mass, but an isotope effect is seen in that the H$_2$ transition is shifted towards lower pressures.
The results are qualitatively similar to results obtained previously in AIMD calculations\cite{lorenzen2010first,Geng2019Thermodynamic} which showed that the proton diffusivity increases rather than decreases with increasing pressure, meaning that the high pressure phase has lower viscosity. 
The diffusion constant remain nonzero in all simulations above the melt line and  a close look at the mean squared displacement for these trajectories (shown in the Supplementary Materials) indicates that even at low temperatures, the simulations are indeed of a flowing liquid. 

\begin{figure}[!htbp]
    \centering
    \includegraphics[width=\columnwidth]{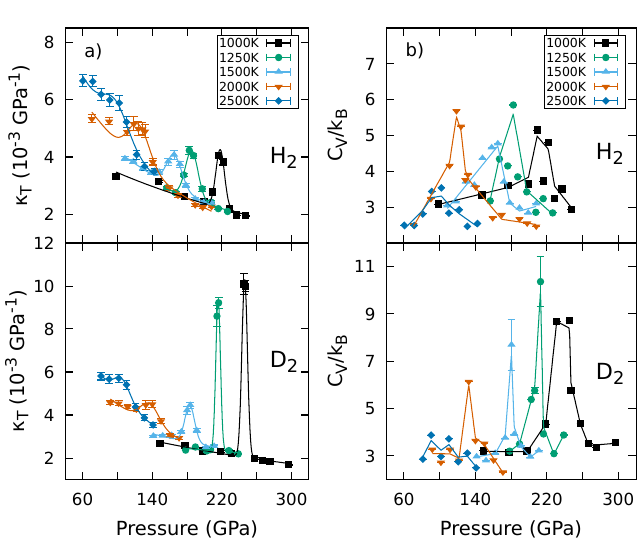}
    \caption{(a) Estimated isothermal compressibility $\kappa_T = -\left(\Delta V/\Delta P \right)_T / V$ calculated using fixed volumes and equilibrated pressures for both isotopes. Solid lines are fits consisting of an exponential plus a Gaussian function. The centre of the Gaussian peak was then used to establish the location of the phase boundary and Widom line. (b) Heat capacities of the full ring polymer systems for H$_2$ and D$_2$ established from fluctuations in energy estimators. Solid lines are a guide to the eye.}
    \label{fig:thermodynamics}
\end{figure}

The LLPT is perhaps best thought of as a chemical reaction between distinct species H and H$_2$. 
This is in part because the transition is anomalous compared to first order transitions usually encountered, as there is no phase separation in the coexistence region: atomic and molecular hydrogen are continually interconverting, and therefore appear miscible\cite{Geng2019Thermodynamic}.
Nevertheless, the location of the phase boundary can be established by considering various thermodynamic quantities. 
Since the system undergoes a small volume change across the isotherm either abruptly or gradually depending on the temperature, this will create a clear signature in the isothermal compressibility $\kappa_T=-\left(\partial V/\partial P \right)_T / V$. 
Here, it is estimated using the equilibrated NVT volumes and pressures using finite differences. 
These results are shown in Fig. \ref{fig:thermodynamics}a), along with a fit to obtain the location of the peak.
In theory, $\kappa_T$ should diverge at a first-order transition and show a peak along an extension of the phase boundary beyond the critical point (the ``Widom line''). While this cannot happen in a finite sized system, the peaks can still be seen to be very sharp for both isotopes at low temperatures. 
For both isotopes the peaks also widen at higher temperatures, which is indicative of a crossover rather than a phase transition.

{}

The other thermodynamic quantity considered is the heat capacity at constant volume $C_V$, as this can be calculated directly from fluctuations in the internal energy. 
This should also diverge at a first-order transition and show a peak along the Widom line, due to the energy required to break the bonds.
To calculate $C_V$ in the PIMD formalism, some modifications must be made to account for interactions between beads. 
For these results, the centroid virial heat capacity is employed (see Supplementary Materials) to calculate the heat capacity of the full PIMD ring polymer system\cite{glaesemann2002improved}.
The results are shown in Fig. \ref{fig:thermodynamics}b) for the various isotherms for both H$_2$ and D$_2$. 
The heat capacity is also clearly peaked like $\kappa_T$, but noisier due to it being a quantity obtained from fluctuations.

In both the quantities considered, the D$_2$ peaks are shifted compared to H$_2$ which further identifies the isotope effect. 
The magnitude of the peaks in both the heat capacity and compressibility is also larger in deuterium than hydrogen at lower temperatures. 
Both the quantities also show broadening of their respective peaks at higher temperatures, clearest in the compressibility, where the transition becomes a crossover rather than a thermodynamic phase transition.



\begin{figure}[!htbp]
    \centering
    \includegraphics[width=0.85\columnwidth]{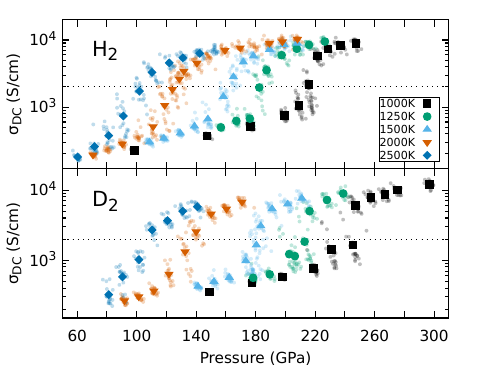}
    \caption{Calculated DC electrical conductivity for H$_2$ and D$_2$ showing the onset of metallization. Scatter points are conductivities of individual samples. A tolerance of 2000 S/cm for metallization is indicated by dotted lines.}
    \label{fig:conductivity}
\end{figure}{}

The prevailing belief about the LLPT is that metallization and dissociation occur together,\cite{mazzola2015distinct,Pierleoni2016Liquid} but in principle there is no reason that these two phenomena must coincide exactly. 
Furthermore, metallization can be observed at a distinctive pressure even beyond the critical point, perhaps by a percolation transition\cite{magdau2017simple}.
Onset of metallization in our simulations was monitored by calculating the Kubo-Greenwood conductivity\cite{calderin2017kubo} from snapshots taken from the PIMD trajectories.  
These results are shown in Fig. \ref{fig:conductivity}, where a minimum conductivity of 2000 S/cm was used to distinguish the metallic phase\cite{weir1996metallization,nellis1999minimum}.
The conductivity steeply increases by two orders of magnitude at the transition pressure, coinciding very closely with dissociation in Fig. \ref{fig:dissociation}: hydrogen metallizes at lower temperature than deuterium. 
Our results show that at all temperatures hydrogen has higher conductivity than deuterium. \
Interestingly, this was noted experimentally by Weir \textit{et al.}\cite{weir1996metallization} but ascribed to a large difference in density, and the exact values remain uncertain\cite{zaghoo2017conductivity}.


\begin{figure*}[!htbp]
    \centering
    \includegraphics[width=0.8\textwidth]{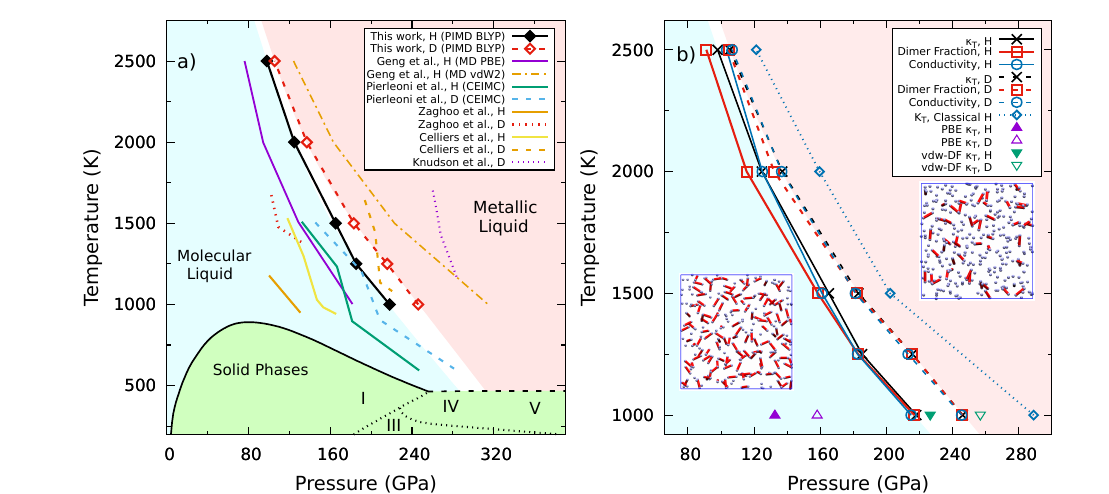}
    \caption{Summary of the results obtained in this work. (a) Location of the phase boundary as established from peaks in the compressibility. The isotope effect is largest in magnitude at lower temperatures. Melting curve\cite{howie2015raman} and the likely continuation of the melting line beyond the I-IV-liquid triple point\cite{dalladay2016evidence} shows where the LLPT phase line is expected to rejoin the solid phase lines. (b) Comparison of the compressibility phase line to the dissociation curve established from peaks in the gradient of the dimer fraction and the  metallization line along with MD runs using classical H nuclei, all using BLYP. Additional NPT runs using PBE and vdw-DF functionals at 1000 K are also shown. Snapshots on either side of the transition are PIMD runs of H$_2$ at 1000 K using BLYP, for 100 GPa and 250 GPa, with bonds highlighted in red.}
    \label{fig:phaseboundary}
\end{figure*}

The results are summarized in Fig. \ref{fig:phaseboundary} on a PT phase diagram, established using peaks in the isothermal compressibility as this was found to give the most distinctive peaks. 
The Clapeyron slope is observed to be negative, which is consistent with the small but negative volume change across the transition.  
The volume change and Clapeyron slope are slightly larger for D$_2$; configurational entropy cannot be calculated precisely, but from volume and slope we estimate it to be about $k_B/2$ per molecule, consistent with the extra degree of freedom from breaking the bond.
The isotope effect can be clearly observed as a shift in the transition pressure, which is largest in magnitude at lower temperatures where NQE effects are most significant. 

The location of the phase boundary is largely influenced by two contributions. Firstly, including NQE significantly lowers the transition pressure. This be seen from complementary AIMD runs using BLYP and classical H nuclei shown in Fig. \ref{fig:phaseboundary}b), where the phase boundary is obtained from the isothermal compressibility using identical methods to the PIMD runs.

Secondly, the location of the LLPT is very sensitive to the choice of exchange-correlation functional, which can shift the whole phase boundary by as much as 100 GPa\cite{Geng2019Thermodynamic}.
This is mainly due to the well-known tendency of the PBE-based functionals\cite{PBE} towards easy metallization compared with functionals which correctly describe the high density-gradient limit\cite{azadi2017role}. Additional PIMD runs, shown in Fig. \ref{fig:phaseboundary}b) (further discussed in the Supplementary Material\cite{SMref}), reveal a similarly significant spread in the location of the phase boundary.

Nevertheless, the isotope shift is found to be far less sensitive to the choice of functional (in contrast to other systems with significant NQE such as liquid water\cite{Pierleoni2016Liquid}), giving a shift of almost 30 GPa at 1000 K.
The magnitude of the isotope effect is smaller than observed in experiments by Zaghoo \textit{et al.}\cite{Zaghoo2018Striking}, but more similar to CEIMC simulations.


The critical point is difficult to locate with MD.  
The discontinuous changes in thermodynamic properties below the critical print are blurred by finite size effects, and anomalous peaks remain along the Widom lines beyond $T_c$. Both the molecular dissociation and metallization curves match up with the thermodynamic phase boundary below the critical point, and thus also exhibit a quantitatively similar isotope effect, but beyond this the dissociation line in particular occurs at slightly lower pressures than the Widom line.


We have shown the existence of a strong isotope effect of several hundred Kelvins in the liquid-liquid transition in hydrogen, with the transformation occurring at lower temperatures and pressures in hydrogen than deuterium, and much lower than in previous work using classical nuclei.  
There is no such effect at the melt line, nor (by definition) in Born-Oppenheimer dynamics. 
So we can confidently ascribe it to the different zero-point energy of hydrogen and deuterium vibrations;  the difference in these $\frac{1}{2}\hbar({\omega_H-\omega_D})$  has the appropriate order of magnitude of hundreds of Kelvin. 

The strong dependence of the calculated LLPT line on the exchange-correlation functional makes it impossible to confidently state the precise location of the LLPT.  
Moreover, above the critical point the transition becomes a crossover and the associated Widom lines depend on the precise quantity being measured.  
In this work, the magnitude of the isotope effect is likely to be more accurate than the exact location of the LLPT line.
While the exact location of the boundary will likely remain contentious until experimental results can pin it down more accurately, \emph{ab initio} PIMD methods are clearly effective in showing that an isotope effect is indeed present the LLPT.  
They also emphasize that experiments on deuterium cannot be used as a proxy for the hydrogen phase diagram.

It is well known the quantum effects are only important at "low" temperatures. 
We show here that low temperature must be interpreted in terms of the relevant quantum energy scales, which for molecular vibration modes can mean shifts in phase boundaries of hundreds of Kelvins at temperatures of thousands of Kelvins.
These conclusions have general applicability to other molecular-dissociation transitions, such as nitrogen and superionic ammonia and water\cite{Celliers2018Insulator,millot2019nanosecond,jiang2018metallization,robinson2020plastic}, meaning that experiments on deuterated samples will significantly overestimate the transition pressures and temperatures compared with the natural material.

\section*{Acknowledgements} We acknowledge the support of the ERC grant HECATE and studentship funding from EPSRC under grant ref EP/L015110/1. This work used the Cirrus UK National Tier-2 HPC Service at EPCC (http://www.cirrus.ac.uk) funded by the University of Edinburgh and EPSRC (EP/P020267/1). We are grateful for computational support from the UK national high performance computing service, ARCHER, for which access was obtained via the UKCP consortium and funded by EPSRC grant ref EP/P022790/1.

\section*{References}
\bibliography{refs_merged}

\end{document}


\title{\textbf{Supplementary materials for\\"Isotope quantum effects in the metallization transition in liquid hydrogen"}}
\author{Sebastiaan van de Bund,$^1$ Heather Wiebe$^1$ \& Graeme J. Ackland$^1$}
\date{}
\maketitle

\begin{enumerate}
    \item \small School of Physics \& Astronomy, The University of Edinburgh, Edinburgh, EH9 3FD, United Kingdom.
\end{enumerate}

\section*{Previous study of the LLPT}

There have been numerous other studies of the LLPT, several of which were published after this manuscript was submitted. Determining the isotope effects was not the primary purpose of any of these papers, and  all of these studies have significant problems which make them unsuitable for this purpose. The most obvious problem is failure to include nuclear quantum effects, which means there is no isotope effect on the phase diagram\cite{morales2010equation,morales2013nuclear,Pierleoni2016Liquid}.  A more subtle problem identified by Cheng \textit{et al.}\cite{cheng2020evidence} is that the molecular liquid equilibrates slowly and freezes in small system sizes. Diffusivity is easily measured in molecular dynamics, and so we disregard studies which do not present this, and are smaller or shorter than the threshold identified by Cheng \textit{et al.} A final problem is that $\Gamma$ point simulations give rise to spurious "polyhydrogen" liquids, a problem which recurs frequently in the literature and preprints \cite{biermann1998proton,biermann1998quantum,geng2012high,geng2015melt,magdau2015charge,magduau2017charge}.

Experimental studies are similarly problematic - there is a large spread in the reported transitions, and above the critical point, the LLPT line is not even uniquely defined.  One cannot sensibly compare results from different types of experiment for hydrogen and deuterium, and hope to extract the isotope effect. Two recent studies\cite{Zaghoo2018Striking,Jiang2020Spectroscopic} have treated both materials with the same approach, but even then dealing with different samples.

The scatter of phase line due to exchange correlation functional is a further problem.  Previous work has shown that BLYP is the most reliable function for metallisation in hydrogen, but even so a large uncertainty remains\cite{clay2014benchmarking,azadi2017role,ackland2015appraisal}.  We do not claim to solve this here, since our focus is on the isotope shift which is due to nuclear quantum effects, rather than the functional.
In Supplementary Table \ref{tab:litcomparison}, we summarise the previous work, demonstrating why it is unsuitable for determining the isotope effect.

\newcommand\T{\rule{0pt}{2.6ex}}       
\newcommand\B{\rule[-1.2ex]{0pt}{0pt}} 

\begin{table}[!tb]
\caption{Table listing major \textit{ab initio} MD studies on the LLPT boundary (MLP = machine-learning potential), comparing the inconsistent range of both simulation parameters and predicted transition pressures. 
Listed transition pressures were calculated at 1500 K. 
Our results provide a comprehensive look at both hydrogen and deuterium with a sufficiently large system size and an appropriate choice of k-points.}
\label{tab:litcomparison}
\resizebox{\textwidth}{!}{%
\begin{tabular}{llccccl}
\hline \hline
\textbf{Authors} & \textbf{XC-functional} & \textbf{Method} & \textbf{H$_2$ P$_T$ (GPa)} & \textbf{D$_2$ P$_T$ (GPa)} & \textbf{System size} & \textbf{k-points} \T \B\\
\hline \hline
\multirow{3}{*}{Lu \textit{et al.}\cite{lu2019towards}} & PBE0+rVV10 & \multirow{3}{*}{PIMD} & 161 & - & \multirow{3}{*}{256} & \multirow{3}{*}{$\Gamma$} \T \\
 & rVV10 &  & 182 & - &  & \\
 & vdw-DF1 &  & 189 & - &  &  \B \\ 
\hline
Hinz \textit{et al.}\cite{hinz2020fully} & SCAN+rVV10 & PIMD (+MD) & $\sim$110 & {$\sim$130} & 256 & $\Gamma$ \T \B \\
\hline
\multirow{2}{*}{Knudson \textit{et al.}\cite{knudson2015direct}} & vdW-DF1 & \multirow{2}{*}{PIMD (+MD)} & - & 203 & \multirow{2}{*}{256} & \multirow{2}{*}{Baldereschi} \T \\
 & vdW-DF2 &  & - & 292  &  &  \B \\
\hline
Tian \textit{et al.}\cite{tian2020first} & PBE & MD & 127 & - & 64 & 3x3x3 \T \B \\
\hline
Lorenzen \textit{et al.}\cite{lorenzen2010first}& PBE & MD & 140 &  - &  512 & Baldereschi  \T \B \\ 
\hline
\multirow{2}{*}{Geng \textit{et al.}\cite{Geng2019Thermodynamic}} & PBE & \multirow{2}{*}{MD}  & 124 & - & \multirow{2}{*}{500-3456} & \multirow{2}{*}{\shortstack{4x4x4 and \\Baldereschi}}      \T \\
 & revPBE-vdW &  & 220 & - &  &  \B \\
\hline
Cheng \textit{et al.}\cite{cheng2020evidence} & PBE, BLYP, QMC & MLP-MD & - & - &  128 (1728)   &  N/A \T \B  \\ 
\hline\hline
\textbf{This work} & \textbf{BLYP} & \textbf{PIMD (+MD)} & \textbf{165} & \textbf{183} & \textbf{360} & \textbf{Baldereschi} \T \B \\ 
\hline \hline
\end{tabular}%
}
\end{table}

\section*{Methods}

The Path Integral technique incorporates NQE by mapping the quantum nucleus onto a set of $\mathcal{P}$ classical replicas, termed "beads", which interact with each other through harmonic springs. Here, classical and Path Integral Molecular Dynamics simulations were performed using the i-Pi\cite{Kapil2019i-PI} universal force engine in conjunction with Quantum Espresso.\cite{QE-2017} PIMD simulations consisted of 180 molecules of H$_2$ or D$_2$ (360 atoms) in a cubic periodic box with an electronic kinetic energy cutoff of 35 Ry and 234 bands. For k-point sampling, the Baldereschi mean value point\cite{Baldereschi1972MeanValue} was chosen due to the tendency of hydrogen to form spurious polymeric structures when sampling is done at the $\Gamma$ point.\cite{magdau2015charge} The widely-used PBE\cite{PBE} functional is known to erroneously favour metallization in hydrogen and so in this work the BLYP functional\cite{blyp-B,blyp-LYP} was used as it has been shown to give a better agreement with experiment for the molecular phases of hydrogen\cite{azadi2013fate}.

Trajectories consisted of an initial NPT equilibration period of 500 steps, followed by 2000 steps in the NVT ensemble which was then used for data analysis. A timestep of 0.5 fs was used in all cases. Temperature was kept constant using the stochastic PILE\_L thermostat\cite{Ceriotti2010Efficient} with a relaxation time of 10 fs, and for the NPT equilibration runs a constant pressure was maintained using the isotropic stochastic barostat\cite{Ceriotti2014i-PI} as implemented in i-PI, with a relaxation time of 100 fs. Five different isotherms were used to locate the LLPT: 1000, 1250, 1500, 2000 and 2500 K. PIMD trajectories were obtained for hydrogen with 12 beads at 1000 and 1250 K, 8 beads at 1500 K and 2000 K and 4 beads at 2500 K. For deuterium 8 beads were used at 1000 and 1250 K, 4 beads at 1500 K, 2000 K and 2500 K. Convergence details are given in the Supplementary Materials below. 

Classical simulations were performed by setting the number of beads to 1. These used identical parameters as above, but only considering four isotherms 1000 K, 1500 K, 2000 K and 2500 K using 1000 steps in the NVT ensemble.

Heat capacities were calculated using the centroid virial estimator, which provides improved convergence over the primitive energy estimator and is given by\cite{glaesemann2002improved}:
\begin{equation}
    \frac{C_V^{CV}}{k_B \beta^2} = \langle \epsilon_T \epsilon_{CV} \rangle - \langle \epsilon_{CV} \rangle^2 - \frac{3N}{2\beta^2},
\end{equation}
for $N$ particles, where $\epsilon_T$ and $\epsilon_{CV}$ are the primitive energy and centroid virial energy estimators respectively. Because of its slower convergence, the primitive energy estimator was shifted by a small constant such that the average of the primitive and centroid virial energy estimators were equal.

\begin{figure}
    \centering
    \includegraphics[width=0.5\textwidth]{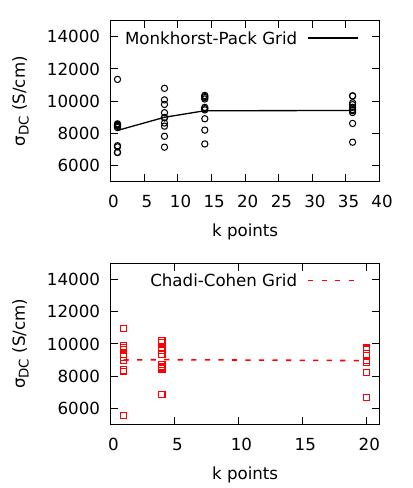}
    \caption{Convergence of k point grid sizes in the SCF calculations used to calculate conductivities.}
    \label{fig:kconv}
\end{figure}

The conductivities were calculated in the Kubo-Greenwood formalism using the KGEC code\cite{calderin2017kubo}. Averages we were calculated using 20 snapshots of each centroid trajectory. A generalisation of the Baldereschi point was used to sample the Brillouin zone, following the approach of Chadi and Cohen\cite{chadi1973special}, to obtain 4 k-points in the SCF calculation. Convergence for 10 independent snapshots at 1000 K and 250 GPa are shown in Fig. \ref{fig:kconv}. The Chadi-Cohen grid is more consistent with previous calculations, and converges the average conductivity faster than a Monkhorst-Pack grid. To evaluate the Kubo-Greenwood conductivity, a width of 1 eV for the Lorentzian used by KGEC was found to give consistent results.
Note that these calculations were performed using a local norm-conserving pseudopotential\cite{hamann1979norm} rather than an ultrasoft pseudopotential\cite{laasonen1993car} in the main trajectories.

\section*{PIMD convergence}

\begin{figure}[!htbp]
    \centering
    \includegraphics[width=0.45\textwidth]{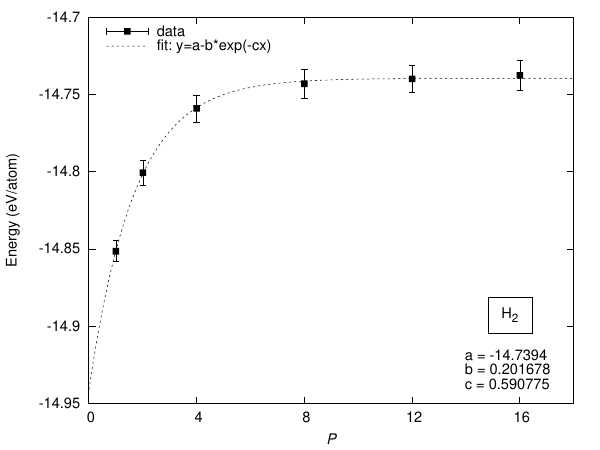}
    \includegraphics[width=0.45\textwidth]{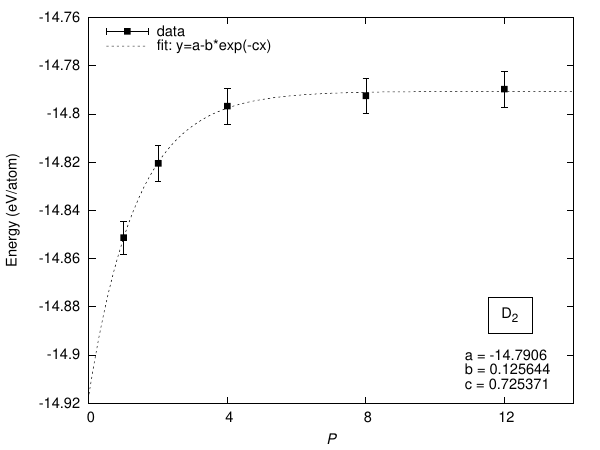}    
    \caption{Convergence of total energy (potential plus kinetic) with respect to number of PIMD beads for liquid hydrogen (left) and liquid deuterium (right) at 100 GPa, 1250 K. The optimal $\mathcal{P}$ was chosen to be at the onset of the plateau, 12 for hydrogen and 8 for deuterium in this case.}
    \label{fig:PIMDconvergence}
\end{figure}

\begin{figure}[!htbp]
    \centering
        \includegraphics[width=0.45\textwidth]{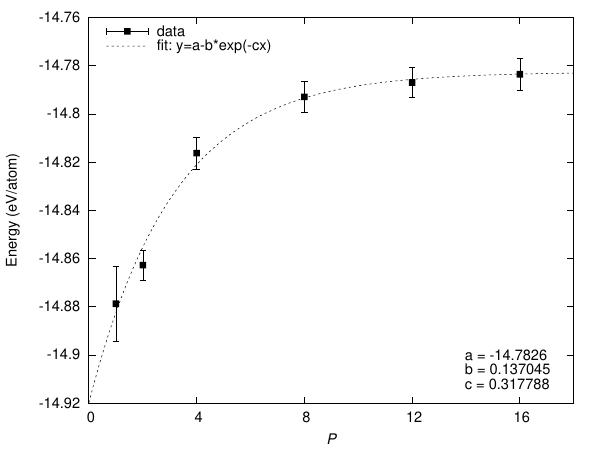}
    \includegraphics[width=0.45\textwidth]{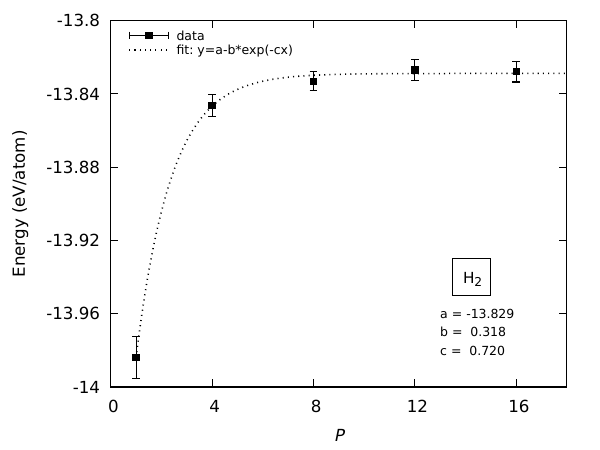}
    \caption{Convergence of total energy (potential plus kinetic) with respect to number of PIMD beads for liquid metallic hydrogen at 100GPa and 250 GPa, 1000 K.}
    \label{fig:PIMDconvergence_highpressure}
\end{figure}

The number of beads $\mathcal{P}$ required to converge NQE is inversely related to temperature, and so we performed convergence tests to determine the optimum $\mathcal{P}$ for each isotherm. The goal was to maintain accuracy of the calculated NQE with a minimum of computational effort. For this, the total energy of the system was calculated as a function of $\mathcal{P}$ at 100 GPa for each temperature, and the optimal $\mathcal{P}$ was chosen as the beginning of the plateau. A sample convergence plot is shown in Fig. \ref{fig:PIMDconvergence}. The optimal $\mathcal{P}$ for hydrogen was chosen to be 12 for 1000 K and 1250 K, 8 for 1500 K and 2000 K, and 4 for 2500 K. For deuterium, where quantum effects are slightly weaker, fewer beads were required for convergence. 8 beads were used for deuterium at 1000 and 1250 K, and 4 beads at 1500 K, 2000 K and 2500 K. 
Note that convergence is slower at lower temperature, but is not significantly affected by pressure, as illustrated in Fig. \ref{fig:PIMDconvergence_highpressure}.

\section*{Radial distribution functions}

The first peak of the radial distribution function (RDF), which corresponds to the covalent H-H bond length, provides an indicator of the amount of H$_2$ molecules present in the system. Here we show sample RDFs for H$_2$ at 1000 K which illustrate the transition between molecular and atomic phases. 

\begin{figure}[!htbp]
    \centering
    \includegraphics[width=0.6\textwidth]{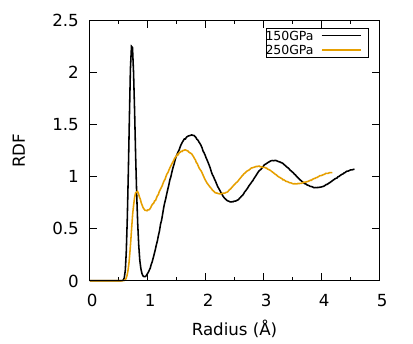}
    \caption{Radial distribution functions for H$_2$ at 1000 K as a function of pressure. The black and blue lines corresponds to the molecular and atomic phases at 150 GPa and 250 GPa, respectively. The yellow dashed line represents the RDF of the solid phase I structure at 150 GPa. The transition between molecular and atomic phases can be clearly seen by the dramatic drop in the first peak of the RDF.}
    \label{fig:rdf}
\end{figure}

\section*{Evidence of liquid phase}

\begin{figure}[!htbp]
    \centering
    \includegraphics[width=0.45\textwidth]{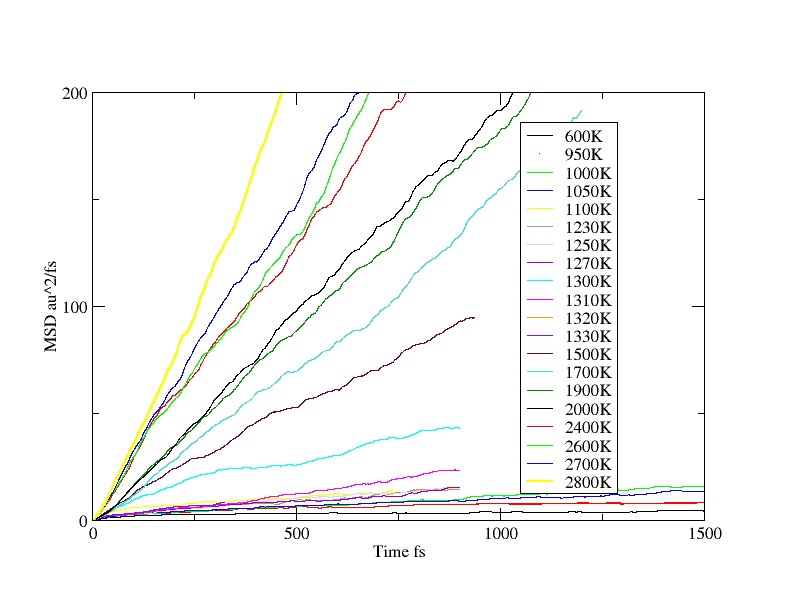}
    \includegraphics[width=0.45\textwidth]{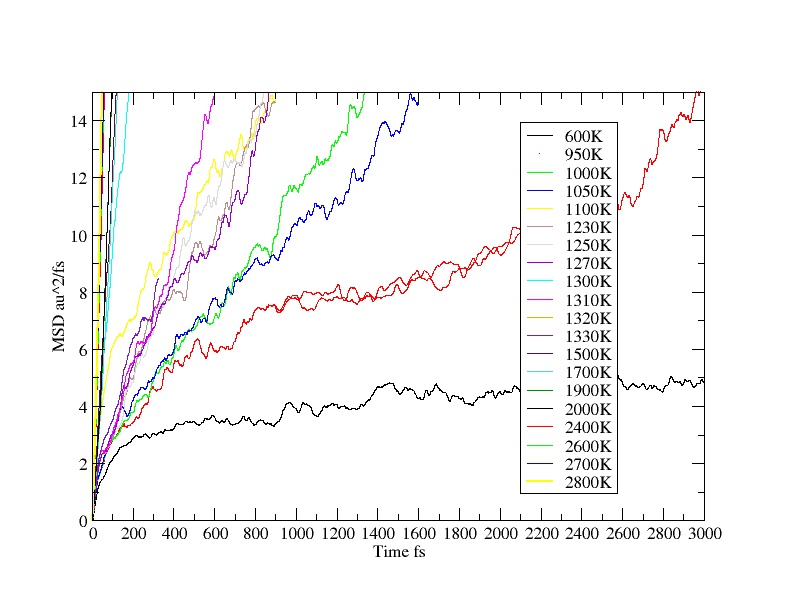}
    \caption{Mean squared displacement from classical-ion simulations 
using the BLYP functional and 256 atoms. MSD is calculated from ensemble averages of all atoms and all start times, $\tau$:  $MSD($t$)=\langle({\bf r}(\tau)-{\bf r}(\tau-t) \rangle$.   These calculations are run in the NVT ensemble at 2.005 \AA$^3$/atom which corresponds to about 140 GPa.
There is an initial steep increase as the molecules do hindered rotation, even in the solid phase.  In the long-time limit, all the runs except 600 K show linear MSD with non-zero slope, and no evidence that diffusion is driven by specific rare events. We conclude that all simulations run above the melt curve are liquid-like.  600K is below the melt curve, but the simulation was initiated from a 1000 K run: the sample appears disordered.  950 K goes through a kind of plateau, which may suggest its getting stuck in a glassy phase: the simulation was restarted from a configuration near the start of this glassy behavior, hence two red lines are shown. Everything near the LLPT region is a good flowing liquid.
}
    \label{fig:MSDclassical}
\end{figure}

\begin{figure}[!htbp]
    \centering
    \includegraphics[width=0.7\textwidth]{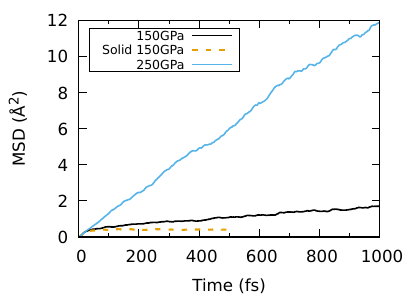}
    \caption{Mean squared displacement from PIMD simulations of H$_2$ at 1000 K. The yellow dashed line represents the solid phase I at 150 GPa, which shows the expected plateau. The black and blue lines represent the molecular phase at 150 GPa, and the atomic phase at 250 GPa. Both curves show a linear MSD in the long-time limit, confirming that these simulations are liquid-like. }
    \label{fig:MSD-PIMD}
\end{figure}

We watched animations of our trajectories to ensure that solidification was not observed. 
This is further  evidenced by the mean squared displacements shown in Fig. \ref{fig:MSDclassical} and \ref{fig:MSD-PIMD}. No solidification, layering, chain formation or other unphysical behaviour was observed in any of our classical or PIMD simulations above the melt line in Fig. 
5.  In the classical simulation at 600 K we observe a non-diffusing, glasslike state which we attribute to cooling below the melt line: this state can not be easily distinguished from the liquid using RDF data alone.
 
\section*{Equation of state}

\begin{figure}[!htbp]
    \centering
    \includegraphics[width=0.7\textwidth]{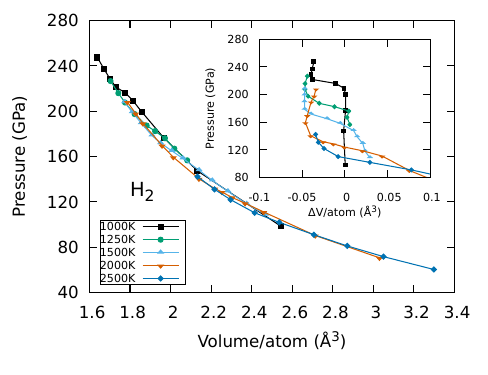}
    \includegraphics[width=0.7\textwidth]{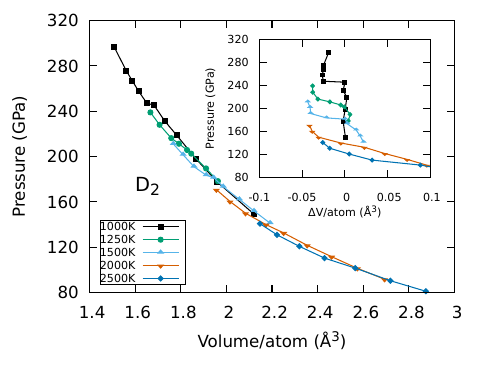}
    \caption{Pressure-volume curves for hydrogen and deuterium. Insets are the same curves with the volume subtracted as obtained from a specific Vinet EoS fit to the molecular low pressure phase at 1000 K as a reference curve in order to emphasize the volume discontinuity at low temperatures.
    }
    \label{fig:EOS}
\end{figure}

Fig. \ref{fig:EOS} shows the PV equation of state for both isotopes. The first order phase transition is evidenced by a plateau or anomaly in the pressure, which is small and difficult to see in the equation of state.   To emphasize it, we also plot the difference from a Vinet equation of state fitted to the molecular phase at 1000 K: this show a distinct step in the volume at 1000K and at 1250 K in deuterium.  The situation is more equivocal for higher temperature, with a distinct change in PV slope across several densities - this suggests a crossover rather than a discontinuous phase transition.

The H$_2$ volume change at 1000 K is $0.025$ \AA$^3$/atom.  From the equation of state we see a Clapeyron slope of approximately 8 K/GPa, from which we can estimate an entropy change around 0.5k$_B$/atom.

\section*{Band gap}
\begin{figure}[!htbp]
    \centering
    \includegraphics[width=0.45\textwidth]{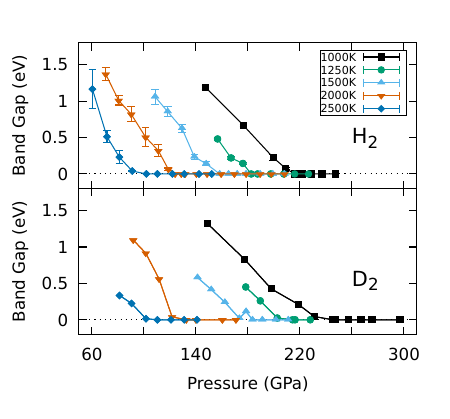}
    \caption{Gap at the Fermi level in the electronic density of states at various pressures, showing the closure of the gap associated with metallization.}
    \label{fig:bandgap}
\end{figure}

In addition to conductivities, band gap closure was also calculated as an additional measure of metallization. These were obtained from the electronic density of states using 20 snapshots of the centroid trajectory. These results are shown in Fig. \ref{fig:bandgap}, showing a similar isotope effect to previous results. They are also in agreement with conductivity calculations presented in the main text.

\section*{Comparison of exchange-correlation functionals}
\begin{figure}
    \centering
    \includegraphics[width=0.45\textwidth]{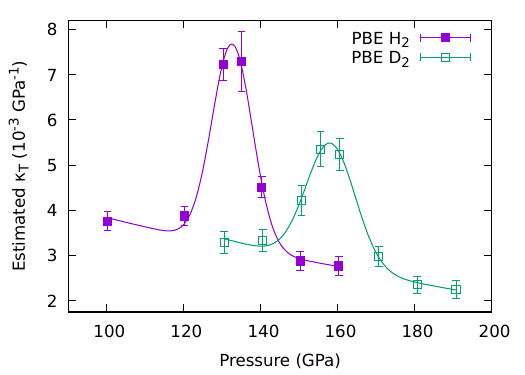}
    \includegraphics[width=0.45\textwidth]{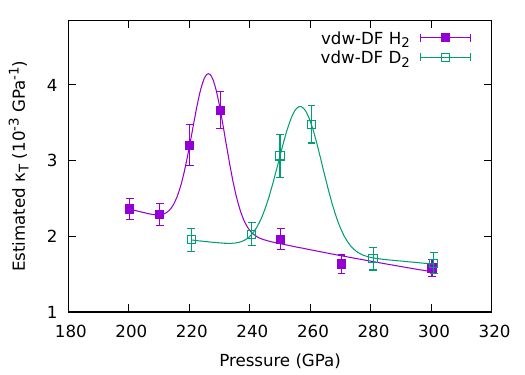}
    \caption{Isothermal compressibility with fits for PBE and vdw-DF exchange-correlation functionals, showing a similar isotopic pressure shift as for BLYP.}
    \label{fig:xc-compress}
\end{figure}

While the choice of exchange-correlation functional has a drastic effect on the location of the LLPT, the isotope effect is not found to be as significantly affected. Fig. \ref{fig:xc-compress} shows the calculated isothermal compressibility of a set of NPT runs of at least 0.5 ps using the PBE\cite{PBE} and vdw-DF\cite{dion2004van} functionals performed at 1000 K, which is where the isotope effect is expected to be strongest. The results are fitted using an exponential background and Gaussian function to obtain the location of the transition using identical methods as for the main results.

The resulting isotope shift on the transition pressures from Fig. \ref{fig:xc-compress} are found to be $\Delta P_\text{PBE} = 26$ GPa and $\Delta P_\text{vdw-DF} = 30$ GPa, compared to the main result of $\Delta P_\text{BLYP} = 27 $ GPa. The spread is thus only a few GPa. Given the shorter length of these simulations and the spacing of the pressures considered, this supports the notion that the isotope effect not only exists even when using other functionals, other choices also yield a similar magnitude. The main effect of changing functional is therefore to significantly shift the location of the phase boundaries in PT space.


This contrasts with the case for other systems where NQE are significant such as liquid H$_2$O. For such a system, the choice of functional can significantly affect the degree to which hydrogen is delocalised, thus creating marked differences in the hydrogen bond network\cite{wang2014quantum}. In contrast, intermolecular interactions in liquid hydrogen are relatively weak: the transition is dominated by local energetics of the two distinct but interconvertible species, hydrogen atoms and hydrogen molecules\cite{Geng2019Thermodynamic}.

The choice of functional will feasibly lead to dissociation energies that differ both from each other, and from the true value. Regardless, these results suggest that any such deviations, including differences in the potential energy surface, will largely cancel out. This leaves the difference in ZPE as the main contribution to the magnitude of the isotopic shift, which is well-described in the PIMD formalism.

\clearpage

\bibliography{refs_merged.bib}